\pdfoutput=1

\documentclass[12pt,a4paper]{article}

\usepackage{ifthen} 
\newboolean{pdflatex}
\setboolean{pdflatex}{true} 

\newboolean{articletitles}
\setboolean{articletitles}{true} 

\newboolean{uprightparticles}
\setboolean{uprightparticles}{false} 

\def\paperauthors{LHCb collaboration} 
\def\paperasciititle{Isospin amplitudes in in Lambda_b to J/psi Lambda(Sigma) and Xi_b to J/psi Xi(Lambda) decays
} 
\def\papertitle{Isospin amplitudes in $\Lb\to\jpsi \PLambda(\PSigma^0)$ and $\PXi_b^0\to\jpsi \PXi^0(\PLambda)$ decays
} 
\def\paperkeywords{{High Energy Physics}, {LHCb}} 
\def\papercopyright{\the\year\ CERN for the benefit of the LHCb collaboration} 
\def\paperlicence{CC-BY-4.0 licence}
\def\paperlicenceurl{https://creativecommons.org/licenses/by/4.0/}

\usepackage[top=1in, bottom=1.25in, left=1in, right=1in]{geometry}

\usepackage{microtype}
\usepackage{lineno}  
\usepackage{xspace} 
\usepackage{caption} 

\usepackage{graphicx}  
\usepackage{color}
\usepackage{colortbl}
\graphicspath{{./figs/}} 
\DeclareGraphicsExtensions{.pdf,.PDF,png,.PNG}

\usepackage{amsmath} 
\usepackage{amssymb}
\usepackage{amsfonts}
\usepackage{upgreek} 

\newcommand*\patchAmsMathEnvironmentForLineno[1]{%
\expandafter\let\csname old#1\expandafter\endcsname\csname #1\endcsname
\expandafter\let\csname oldend#1\expandafter\endcsname\csname
end#1\endcsname
 \renewenvironment{#1}%
   {\linenomath\csname old#1\endcsname}%
   {\csname oldend#1\endcsname\endlinenomath}%
}
\newcommand*\patchBothAmsMathEnvironmentsForLineno[1]{%
  \patchAmsMathEnvironmentForLineno{#1}%
  \patchAmsMathEnvironmentForLineno{#1*}%
}
\AtBeginDocument{%
\patchBothAmsMathEnvironmentsForLineno{equation}%
\patchBothAmsMathEnvironmentsForLineno{align}%
\patchBothAmsMathEnvironmentsForLineno{flalign}%
\patchBothAmsMathEnvironmentsForLineno{alignat}%
\patchBothAmsMathEnvironmentsForLineno{gather}%
\patchBothAmsMathEnvironmentsForLineno{multline}%
\patchBothAmsMathEnvironmentsForLineno{eqnarray}%
}

\usepackage{hyperxmp}

\usepackage[pdftex,
            pdfauthor={\paperauthors},
            pdftitle={\paperasciititle},
            pdfkeywords={\paperkeywords},
            pdfcopyright={Copyright (C) \papercopyright},
            pdflicenseurl={\paperlicenceurl}]{hyperref}

\usepackage[all]{hypcap} 

\usepackage{xspace} 
\usepackage{upgreek}

\def\lhcb   {\mbox{LHCb}\xspace}

\def\MagUp {\mbox{\em Mag\kern -0.05em Up}\xspace}

\ifthenelse{\boolean{uprightparticles}}%
{

 \def\Pmu         {\ensuremath{\upmu}\xspace}

 \def\Ppi         {\ensuremath{\uppi}\xspace}

 \def\Ppsi        {\ensuremath{\uppsi}\xspace}

 \def\PDelta      {\ensuremath{\Delta}\xspace}                 
 \def\PXi         {\ensuremath{\Xi}\xspace}                 
 \def\PLambda     {\ensuremath{\Lambda}\xspace}                 
 \def\PSigma      {\ensuremath{\Sigma}\xspace}                 
 \def\POmega      {\ensuremath{\Omega}\xspace}                 
 \def\PUpsilon    {\ensuremath{\Upsilon}\xspace}

 \def\PB      {\ensuremath{\mathrm{B}}\xspace}                 
                  
 \def\PD      {\ensuremath{\mathrm{D}}\xspace}

 \def\PJ      {\ensuremath{\mathrm{J}}\xspace}                 
 \def\PK      {\ensuremath{\mathrm{K}}\xspace}

 \def\Pb      {\ensuremath{\mathrm{b}}\xspace}

 \def\Pi      {\ensuremath{\mathrm{i}}\xspace}

 \def\Ps      {\ensuremath{\mathrm{s}}\xspace}

 \def\thebaroffset{0.0em}
}
{

 \def\Pmu         {\ensuremath{\mu}\xspace}

 \def\Ppi         {\ensuremath{\pi}\xspace}

 \def\Ppsi        {\ensuremath{\psi}\xspace}                 
                  
 \mathchardef\PDelta="7101
 \mathchardef\PXi="7104
 \mathchardef\PLambda="7103
 \mathchardef\PSigma="7106
 \mathchardef\POmega="710A
 \mathchardef\PUpsilon="7107
                  
 \def\PB      {\ensuremath{B}\xspace}                 
                  
 \def\PD      {\ensuremath{D}\xspace}

 \def\PJ      {\ensuremath{J}\xspace}                 
 \def\PK      {\ensuremath{K}\xspace}

 \def\Pb      {\ensuremath{b}\xspace}

 \def\Pi      {\ensuremath{i}\xspace}

 \def\Ps      {\ensuremath{s}\xspace}

 \def\thebaroffset{0.18em}
}
\newcommand{\offsetoverline}[2][\thebaroffset]{\kern #1\overline{\kern -#1 #2}}%

\makeatletter
\ifcase \@ptsize \relax
  \newcommand{\miniscule}{\@setfontsize\miniscule{4}{5}}
\or
  \newcommand{\miniscule}{\@setfontsize\miniscule{5}{6}}
\or
  \newcommand{\miniscule}{\@setfontsize\miniscule{5}{6}}
\fi
\makeatother

\DeclareRobustCommand{\optbar}[1]{\shortstack{{\miniscule (\rule[.5ex]{1.25em}{.18mm})}
  \\ [-.7ex] $#1$}}

\def\mumu       {{\ensuremath{\Pmu^+\Pmu^-}}\xspace}

\def\squark    {{\ensuremath{\Ps}}\xspace}

\def\bquark    {{\ensuremath{\Pb}}\xspace}

\def\pion   {{\ensuremath{\Ppi}}\xspace}

\def\pip    {{\ensuremath{\pion^+}}\xspace}
\def\pim    {{\ensuremath{\pion^-}}\xspace}

\def\kaon    {{\ensuremath{\PK}}\xspace}

\def\KorKbar {\kern \thebaroffset\optbar{\kern -\thebaroffset \PK}{}\xspace}

\def\KS      {{\ensuremath{\kaon^0_{\mathrm{S}}}}\xspace}

\def\D       {{\ensuremath{\PD}}\xspace}

\def\DorDbar {\kern \thebaroffset\optbar{\kern -\thebaroffset \PD}\xspace}

\def\Dp      {{\ensuremath{\D^+}}\xspace}
\def\Dm      {{\ensuremath{\D^-}}\xspace}

\def\DpDm    {\ensuremath{\Dp {\kern -0.16em \Dm}}\xspace}

\def\B       {{\ensuremath{\PB}}\xspace}
\def\Bbar    {{\ensuremath{\offsetoverline{\PB}}}\xspace}

\def\BorBbar {\kern \thebaroffset\optbar{\kern -\thebaroffset \PB}\xspace}

\def\Bzb     {{\ensuremath{\Bbar{}^0}}\xspace}
\def\Bd      {{\ensuremath{\B^0}}\xspace}

\def\BdorBdbar {\kern \thebaroffset\optbar{\kern -\thebaroffset \Bd}\xspace}

\def\Bs      {{\ensuremath{\B^0_\squark}}\xspace}

\def\BsorBsbar {\kern \thebaroffset\optbar{\kern -\thebaroffset \Bs}\xspace}

\def\jpsi     {{\ensuremath{{\PJ\mskip -3mu/\mskip -2mu\Ppsi}}}\xspace}

\def\Y#1S{\ensuremath{\PUpsilon{(#1S)}}\xspace}

\def\Lz          {{\ensuremath{\PLambda}}\xspace}

\def\LorLbar     {\kern \thebaroffset\optbar{\kern -\thebaroffset \PLambda}\xspace}

\def\Xires       {{\ensuremath{\PXi}}\xspace}

\def\Lb           {{\ensuremath{\Lz^0_\bquark}}\xspace}

\def\Xib          {{\ensuremath{\Xires_\bquark}}\xspace}

\newcommand{\decay}[2]{\ensuremath{#1\!\to #2}\xspace} 

\def\to                 {\ensuremath{\rightarrow}\xspace}

\def\CP                {{\ensuremath{C\!P}}\xspace}

\def\AT#1     {\ensuremath{A_{\mathrm{T}}^{#1}}\xspace}           

\def\C#1      {\ensuremath{\mathcal{C}_{#1}}\xspace}                       
\def\Cp#1     {\ensuremath{\mathcal{C}_{#1}^{'}}\xspace}                    
\def\Ceff#1   {\ensuremath{\mathcal{C}_{#1}^{\mathrm{(eff)}}}\xspace}        
\def\Cpeff#1  {\ensuremath{\mathcal{C}_{#1}^{'\mathrm{(eff)}}}\xspace}       
\def\Ope#1    {\ensuremath{\mathcal{O}_{#1}}\xspace}                       
\def\Opep#1   {\ensuremath{\mathcal{O}_{#1}^{'}}\xspace}                    

\newcommand{\aunit}[1]{\ensuremath{\text{\,#1}}}       

\newcommand{\tev}{\aunit{Te\kern -0.1em V}\xspace}
\newcommand{\gev}{\aunit{Ge\kern -0.1em V}\xspace}
\newcommand{\mev}{\aunit{Me\kern -0.1em V}\xspace}
\newcommand{\kev}{\aunit{ke\kern -0.1em V}\xspace}
\newcommand{\ev}{\aunit{e\kern -0.1em V}\xspace}
 
\newcommand{\mevc}{\ensuremath{\aunit{Me\kern -0.1em V\!/}c}\xspace}
\newcommand{\gevc}{\ensuremath{\aunit{Ge\kern -0.1em V\!/}c}\xspace}
\newcommand{\mevcc}{\ensuremath{\aunit{Me\kern -0.1em V\!/}c^2}\xspace}
\newcommand{\gevcc}{\ensuremath{\aunit{Ge\kern -0.1em V\!/}c^2}\xspace}

\def\fb   {\ensuremath{\aunit{fb}}\xspace}
\def\invfb   {\ensuremath{\fb^{-1}}\xspace}

\def\ps   {\ensuremath{\aunit{ps}}\xspace}

\def\gsim{{~\raise.15em\hbox{$>$}\kern-.85em
          \lower.35em\hbox{$\sim$}~}\xspace}
\def\lsim{{~\raise.15em\hbox{$<$}\kern-.85em
          \lower.35em\hbox{$\sim$}~}\xspace}

\def\pt         {\ensuremath{p_{\mathrm{T}}}\xspace}

\def\evtgen     {\mbox{\textsc{EvtGen}}\xspace}

\def\geant      {\mbox{\textsc{Geant4}}\xspace}

\def\photos     {\mbox{\textsc{Photos}}\xspace}

\def\pythia     {\mbox{\textsc{Pythia}}\xspace}

\def\tell1  {TELL1\xspace}
\def\ukl1   {UKL1\xspace}

\usepackage{cite} 
\usepackage{mciteplus}

\usepackage{longtable} 

\begin{document}

\renewcommand{\thefootnote}{\fnsymbol{footnote}}
\setcounter{footnote}{1}


\begin{titlepage}
\pagenumbering{roman}

\vspace*{-1.5cm}
\centerline{\large EUROPEAN ORGANIZATION FOR NUCLEAR RESEARCH (CERN)}
\vspace*{1.5cm}
\noindent
\begin{tabular*}{\linewidth}{lc@{\extracolsep{\fill}}r@{\extracolsep{0pt}}}
\ifthenelse{\boolean{pdflatex}}
{\vspace*{-1.5cm}\mbox{\!\!\!\includegraphics[width=.14\textwidth]{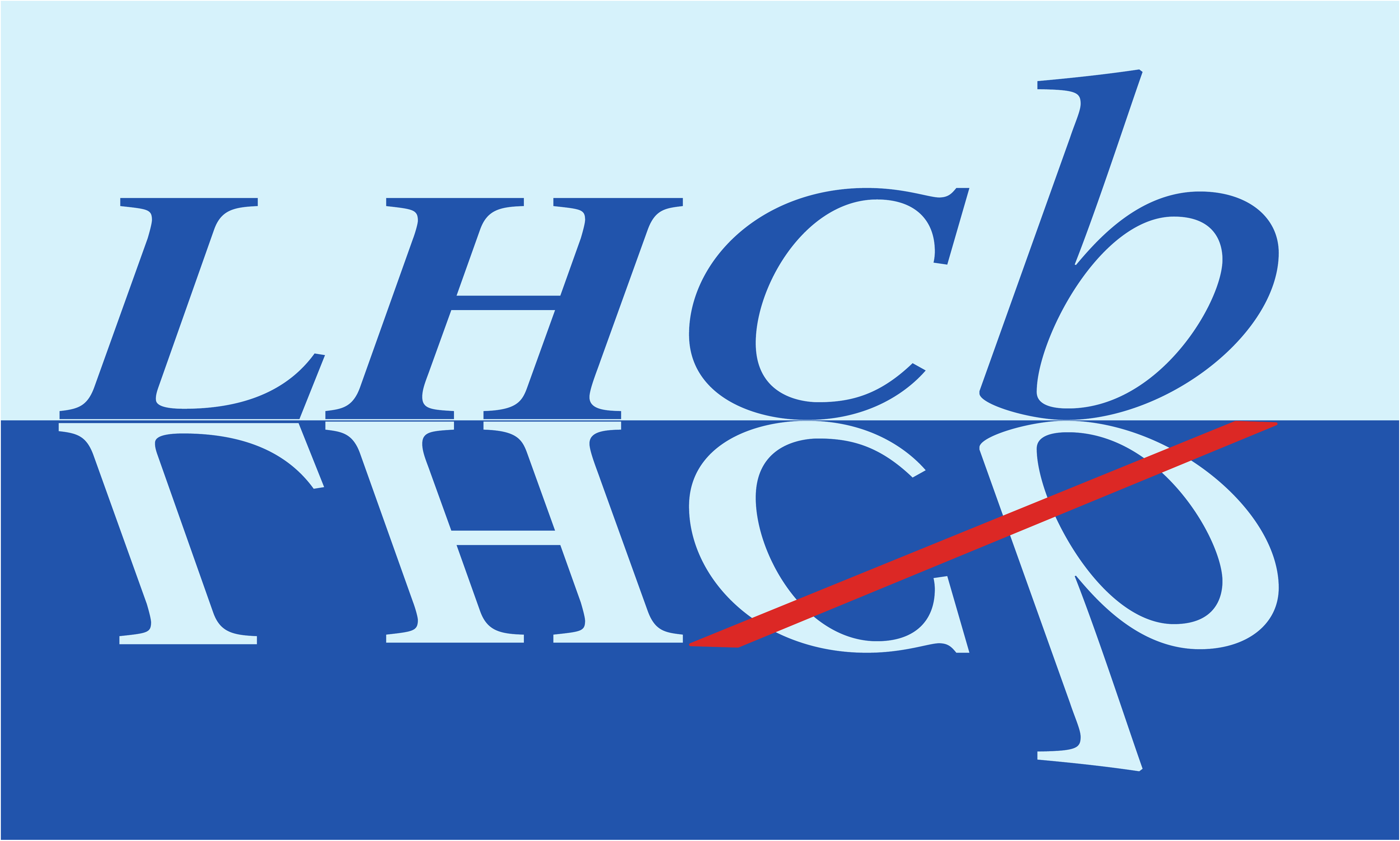}} & &}%
{\vspace*{-1.2cm}\mbox{\!\!\!\includegraphics[width=.12\textwidth]{lhcb-logo.eps}} & &}%
\\
 & & CERN-EP-2019-268 \\  
 & & LHCb-PAPER-2019-039 \\  
 & & \today \\ 
 & & \\
\end{tabular*}

\vspace*{3.0cm}

{\normalfont\bfseries\boldmath\huge
\begin{center}
  \papertitle
\end{center}
}

\vspace*{2.0cm}

\begin{center}
\paperauthors\footnote{Authors are listed at the end of this paper.}
\end{center}

\vspace{\fill}

\begin{abstract}
  \noindent
Ratios of isospin amplitudes in hadron decays are a 
useful probe of the interplay between weak and strong interactions, and allow searches for physics beyond the Standard Model.
We present the first results on isospin amplitudes in $b$-baryon decays, using data corresponding 
to an integrated luminosity of 8.5\invfb, collected with the LHCb detector in $pp$ collisions 
at center of mass energies of 7, 8 and 13 TeV.
The isospin amplitude ratio $|A_1(\Lb\to\jpsi \PSigma^0)/A_0(\Lb\to\jpsi\PLambda)|$, where the subscript on $A$ indicates the final-state isospin, is measured to be less than $1/21.8 $ at 95\% confidence level.
The Cabibbo suppressed $\PXi_b^0\to\jpsi\PLambda$ decay is observed for the first time, allowing for the measurement $|A_0(\PXi_b^0\to\jpsi\PLambda)/A_{1/2}(\PXi_b^0\to\jpsi\PXi^0)| =0.37 \pm 0.06\pm 0.02$, where the uncertainties are statistical and systematic, respectively.
 \end{abstract}

\vspace*{2.0cm}

\begin{center}
  Published in Physical Review Letters 124 (2020) 111802
  \end{center}
\vspace{\fill}

{\footnotesize
\centerline{\copyright~\papercopyright. \href{\paperlicenceurl}{\paperlicence}.}}
\vspace*{2mm}

\end{titlepage}


\newpage
\setcounter{page}{2}
\mbox{~}
%
%
%
%

\cleardoublepage


\renewcommand{\thefootnote}{\arabic{footnote}}
\setcounter{footnote}{0}



\pagestyle{plain} 
\setcounter{page}{1}
\pagenumbering{arabic}


%

Measurements of ratios of isospin amplitudes $A_i$ ($i$ denotes the final state isospin) in hadronic weak decays are a sensitive way to probe the interplay between strong and weak interactions. Such ratios can also reveal the presence of non-Standard Model amplitudes.
For example, in $K\to\pi\pi$ decays the experimentally determined ratio $|A_0/A_2| \approx 22.5$ has not been understood for over 50 years \cite{Cheng:1988va}.
Recent models of the strong dynamics  \cite{Buras:2014maa} and lattice gauge calculations \cite{Boyle:2012ys,*Blum:2015ywa,*Ishizuka:2015oja} for these decays give only partial explanations.
Determinations of isospin amplitudes from  $D\to\pi\pi$ and $B\to \pi\pi$ decays, using input from other two-body decays into light hadrons, found $|A_0/A_2|\approx 2.5$  \cite{Franco:2012ck}, and $|A_0/A_2|\approx 1.0$  \cite{Grinstein:2014aza}, respectively.

\begin{figure}[b]
\vskip -0.35cm
\centering
\includegraphics[width=0.45\textwidth]{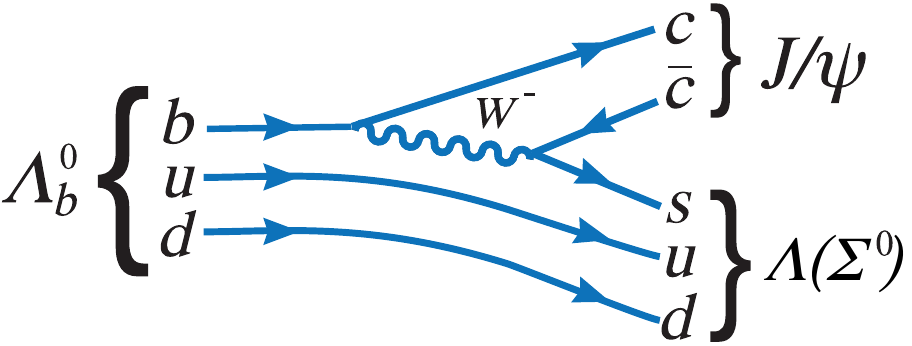}\hspace{4mm}\includegraphics[width=0.45\textwidth]{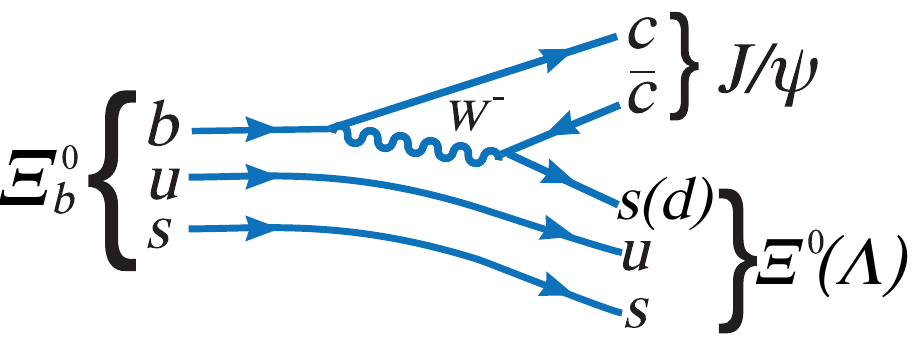}
\caption{\small Leading order Feynman diagrams for $\Lb\to\jpsi \PLambda(\PSigma^0)$ and $\PXi_b^0\to\jpsi \PXi^0(\PLambda)$ decays.}
\label{feynJLam}
\end{figure}

In this Letter, we investigate $\Lb\to\jpsi \PLambda(\PSigma^0)$ and  $\PXi_b^0\to\jpsi \PXi^0(\PLambda)$ decays. (Mention of a specific decay implies the use of its charge-conjugate as well.) The leading order Feynman diagrams for all four processes are shown in Fig.~\ref{feynJLam}. The isospins of the $\jpsi$ meson and $\PLambda$ baryon are zero, and that of the $\PSigma^0$ baryon is one. The isospin of the \Lb baryon is predicted by the quark model to be zero. Since the $b\to c\overline{c} s$ weak operator involves no isospin change, if this prediction is correct, we expect a dominant $A_0$ amplitude and a preference for the $\jpsi\PLambda$ final state over $\jpsi\PSigma^0$, which proceeds via the $A_1$ amplitude. Isospin breaking effects are possible due to the difference in mass and charge of the $u$ and $d$ quarks and can also be induced by QED, electroweak-penguin, or new physics processes \cite{Grossman:1999av}. If the \Lb baryon comprises a $ud$ diquark such effects should be small. Mixing of the $\PLambda$ and $\PSigma^0$ baryons is also predicted to be small, $\sim \!1^{\circ}$, and could contribute $\sim\! 0.01$ to the $|A_1/A_0|$ amplitude ratio \cite{Coleman:1961jn,*Dalitz:1964es,*Kordov:2019oer}.  A severely suppressed $\jpsi\PSigma^0$ final state would determine the isospin of the $\Lb$ baryon to be zero.
Some previous LHCb analyses of $\Lb$ decays made assumptions concerning isospin amplitudes. For instance, the pentaquark analysis, using the $\Lb\to\jpsi K^- p$ channel \cite{Aaij:2015tga}, assumed that the $A_0$ amplitude was dominant, and in the measurement of $|V_{ub}/V_{cb}|$ using $\Lb\to p\mu^-\overline{\nu}$ decays \cite{Aaij:2015bfa} the $A_{3/2}$ amplitude was assumed to be much smaller than the $A_{1/2}$ amplitude.

In $\PXi_b^0\to\jpsi \PXi^0(\PLambda)$ decays, taking the $\PXi_b$ isospin as 1/2, the final state results from an isospin change of zero (1/2) and has $A_i=A_{1/2}~(A_0)$. In the reaction resulting in a final state $\PLambda$ baryon,
the weak transition changes isospin due to the $b\to c \overline{c}d$ rather than the $b\to c \overline{c}s$ transition. Here we investigate if the larger isospin change is suppressed, or if the decay amplitude is independent of the isospin change.  Note that we measure the decay $\PXi_b^-\to\jpsi \PXi^-$ for two purposes: as a proxy for $\PXi_b^0\to\jpsi \PXi^0$, which is difficult for us to measure, and to determine the background in $\jpsi\PLambda$ mass spectrum from these decays where $\PXi\to\PLambda\pi$.

 The LHCb detector is a
single-arm forward spectrometer covering the pseudorapidity range
$2 < \eta < 5$,
described in detail in Refs.~\cite{LHCb-DP-2008-001,LHCb-DP-2014-002}.
The trigger~\cite{LHCb-DP-2012-004}  consists of a
hardware stage, based on information from the calorimeter and muon
systems, followed by a software stage, which reconstructs charged particles. {Natural units are used here with $c=\hbar=1$.}
We use data collected by the LHCb detector, corresponding to  $1.0~\rm fb^{-1}$ of integrated luminosity in 7\tev $pp$  collisions, $2.0~\rm fb^{-1}$ at 8\tev , and $5.5~\rm fb^{-1}$ collected at 13\tev. Hereafter, the data recorded at 7 and 8\tev is referred to as Run 1 and the data recorded at 13\tev is referred to as Run 2.

Simulation is required to model the effects of the detector acceptance and selection requirements.
  We generate $pp$ collisions using
\pythia~\cite{Sjostrand:2007gs,*Sjostrand:2006za}
 with a specific \lhcb configuration~\cite{LHCb-PROC-2010-056}.  Decays of unstable particles
are described by \evtgen~\cite{Lange:2001uf}, where final-state
radiation is generated using \photos~\cite{Golonka:2005pn}. The
interaction of the particles with the detector, and its response,
are implemented using the \geant
toolkit~\cite{Allison:2006ve, *Agostinelli:2002hh} as described in
Ref.~\cite{LHCb-PROC-2011-006}. The lifetimes for the \Lb and $\PXi_b^-$ baryons are taken as 1.473 and 1.572\ps \cite{PDG}, respectively. All simulations are performed separately for Run 1 and Run 2.

Our strategy is to fully reconstruct the $\jpsi\PLambda$ final state and partially reconstruct the $\jpsi\PSigma^0$ mode by ignoring the photon from the $\PSigma^0\to\gamma\PLambda$ decay, because of the low efficiency of the calorimeter at small photon energies. For these decays the $\jpsi\PLambda$ mass distribution is almost uniform in the mass range 5350--5620\mev. We simulate its shape and then fit the mass distribution to ascertain its size. 
The \jpsi meson is reconstructed through the $\jpsi\to\mumu$ decay.
Candidates are formed by combining two oppositely charged tracks identified as muons, with transverse momentum $\pt>550$~\!MeV. Each of the two muons are required to have a maximal $\chi^{2}$ of distance of closest approach of 30 and are also required to form a vertex with $\chi^2_{\rm vtx}<16$. The $\jpsi$ candidate is required to have a decay length significance from every primary vertex, PV, of greater than 3 and a mass in the range 3049--3140~\!MeV.

Candidate $\PLambda$ baryons  are formed from a pair of identified proton and $\pi^-$ particles, each with momentum greater than 2~\!GeV.  Due to their long lifetime and high boost, a majority of the $\PLambda$ baryons decay after the
vertex detector. However, we use only putative decays that occur inside the vertex detector. Each of the two tracks must be inconsistent with having originated from a PV, have a maximal $\chi^{2}$ of distance of closest approach of 30, form a vertex with $\chi^2_{\rm vtx}<12$ that is separated from that PV by more than 3 standard deviations, and have a mass between 1105 and 1124~\!MeV. In addition, we eliminate candidates that when interpreted as $\pi^+\pi^-$ fall within 7.5\mev of the known \KS mass.
Candidate $\PXi^-\to\PLambda\pi^-$ decays are reconstructed using the criteria in Ref.~\cite{Aaij:2019ezy}, with the additional requirement that the $\PXi^-$ decays in the LHCb vertex detector. These are combined with selected $\jpsi$ mesons to form candidate $\PXi^-_b$ baryons.

We improve the $\jpsi\PLambda$ mass resolution by constraining the $\jpsi$ and $\PLambda$ candidates to their known masses and their decay products to originate from each of the relevant decay vertices;  we also constrain the $\jpsi$ and the $\PLambda$ candidates to come from the same decay point \cite{Hulsbergen:2005pu}.

After these selections, we use two
boosted decision trees~(BDT)~\cite{Breiman,AdaBoost} implemented in the TMVA
toolkit~\cite{Hocker:2007ht,*Stelzer:2008zz} to further separate signal from background.
The first BDT is trained to reject generic $b\to \jpsi X$ decays where $X$ contains one or more charged tracks.
We train this ``isolation" BDT using the following information: the $\chi^2_{\rm IP}$ of additional charged tracks with respect to the $\jpsi$ vertex, where $\chi^2_{\rm IP}$ is defined as the difference in the $\chi^2_{\rm vtx}$ of the \jpsi vertex reconstructed with and
without the track being considered; the $\chi^2_{\rm vtx}$ of the vertex formed by the $\jpsi$ plus each additional track; the minimum $\chi^2_{\rm IP}$ of the additional track with respect to any PV; and the \pt of the additional track.  For the isolation BDT training, we use samples of $\Lb\to\jpsi\PLambda$ and $B^-\to\jpsi K^-$ candidates for the signal and background models, respectively. Both samples are background subtracted using the $sPlot$ technique \cite{Pivk:2004ty}.  The output of the isolation BDT is used as an input variable in the final BDT.

The twenty discrimination variables used in the final BDT are listed in the Supplemental material. These mostly exploit the topology of the decay using the vertexing properties of the $\jpsi$, $\PLambda$, and $\Lb$ candidates, and particle identification of their decay products.
The signal sample again is background-subtracted $\Lb\to\jpsi\PLambda$ combinations.
For background training we use candidates in the upper sideband with $\jpsi\PLambda$ masses between $5.7-6.0\gev$, excluding events in  $5.77-5.81\gev$ to avoid including $\PXi_b^0 \to \jpsi\PLambda$ decays in the background sample. We use k-folding cross validation with five folds in both BDTs, to avoid any possible bias \cite{Bagoly:2017bvc}.
The final BDT selection is optimized to maximize the Punzi figure of merit, $\epsilon_s/(\sqrt{B}+1.5)$ \cite{Punzi:2003bu}, where $\epsilon_s$ is the efficiency of the final BDT selection on simulated $\Lb\to\jpsi\PSigma^0$ decays and $B$ is the number of background candidates in the above defined sideband that pass the BDT requirement, scaled to the width of the $\jpsi\PSigma^0$ signal window.
The analysis is performed separately on Run 1 and Run 2 data. The resulting $\jpsi\PLambda$ mass spectrum for Run 2 data is shown in Fig.~\ref{LOG_RUN2_COLOR}. The Run 1 mass distribution is similar and is shown in the Supplemental material.

\begin{figure}[t]
\centering
\includegraphics[width=0.9\textwidth]{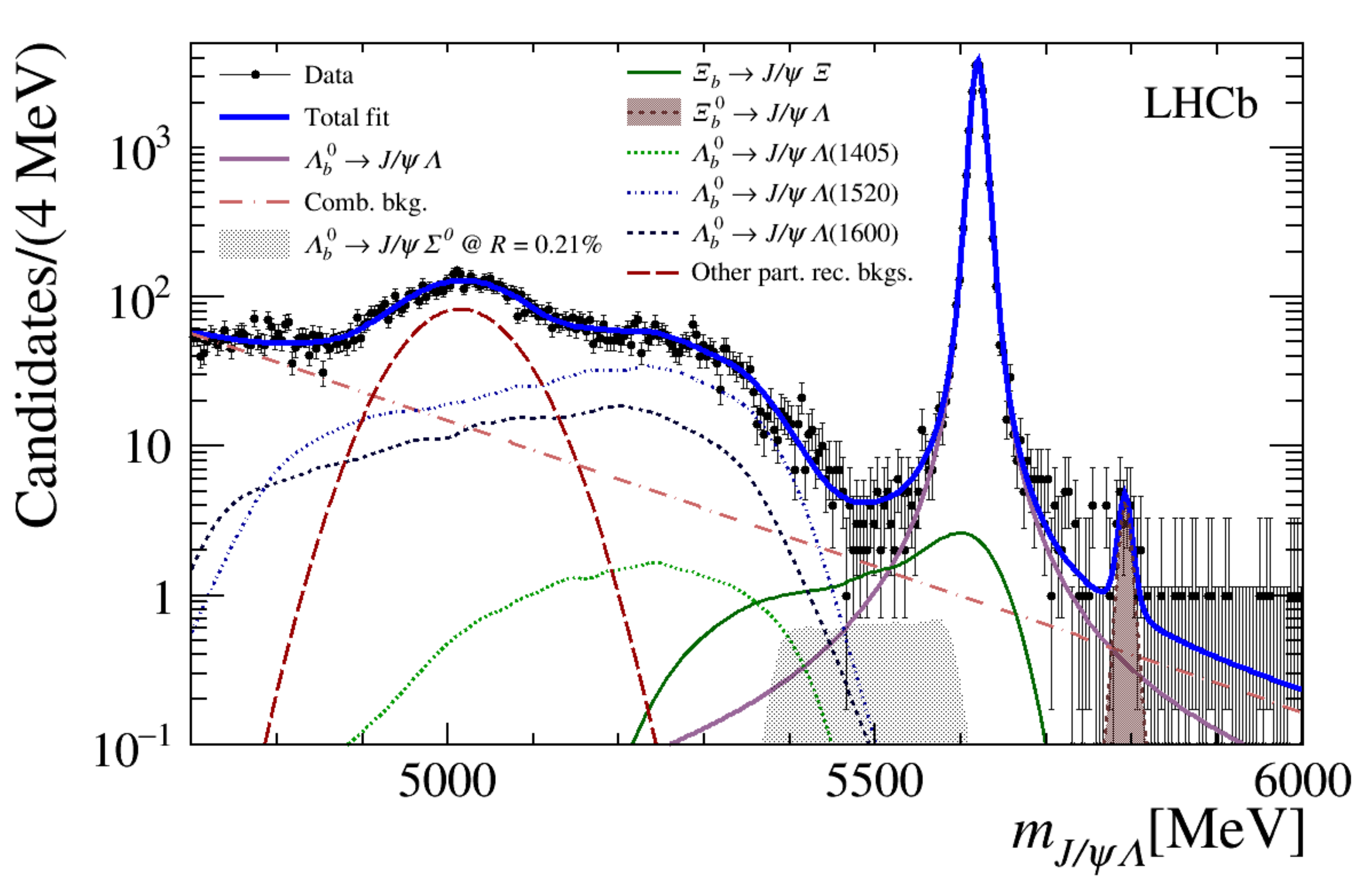}%
\caption{\small
Distribution of the $\jpsi\PLambda$ mass for Run 2 data. Error bars without data points indicate empty bins. Also shown is the projection of the joint fit to the data. The thick (blue) solid curve shows the total fit. For illustrative purposes, the $\Lb\to\jpsi\PSigma^0$ signal component is artificially scaled to its measured upper limit. The shapes are identified in the legend. }
\label{LOG_RUN2_COLOR}
\end{figure}

There are two signal peaks evident in the mass distribution in Fig.~\ref{LOG_RUN2_COLOR}. The larger is due to $\Lb\to\jpsi\PLambda$ decays, and the smaller corresponds to $\PXi_b^0\to\jpsi\PLambda$ decays. The latter is a heretofore unobserved Cabibbo-suppressed decay.
The Run 1 and 2 mass distribution data are fit jointly to determine the $\Lb\to\jpsi\PLambda$, $\Lb\to\jpsi\PSigma^0$ and $\PXi_b^0\to\jpsi\PLambda$ yields. The $\Lb\to\jpsi\PSigma^0$ signal is modeled using a Gaussian kernel \cite{Cranmer:2000du} shape fit to simulation. The $\Lb\to\jpsi\PLambda$ signal is described by a Hypatia function, whose tail parameters are fixed from simulation, with the mass and width allowed to vary in the fit to the data. \cite{Santos:2013gra}. The $\PXi_b^0\to\jpsi\PLambda$ peak is fit to the same shape but with its mean constrained to the fitted $\Lb$ mass plus the known $\PXi_b^0-\Lb$ mass difference of $172.5\mev$ \cite{PDG}.

While most of the candidates above the $\Lb$ peak are the result of combinatoric background, those below are due to additional sources. One is due to $\Lb\to\jpsi \PLambda^*$ decays, with $\PLambda^*\to \PSigma^0\pi^0$ and $\PSigma^0\to \gamma\PLambda$. Here, $\PLambda^*$ denotes strange-baryon resonances ranging from 1405 MeV to 2350 MeV in mass. Another source comprises partially reconstructed $\Lb\to\psi(2S)\PLambda$ decays, where $\psi(2S)\to\pi\pi\jpsi$. These decays mainly populate masses lower than the $\Lb\to\jpsi\PSigma^0$ signal, but need to be included to accurately model the combinatoric background.
The existence of the $\Lb\to\jpsi \PLambda^*$ channels was demonstrated in a study of $\Lb\to\jpsi K^-p$ decays \cite{Aaij:2015tga}. We can model the resulting $\jpsi\PLambda$ mass shapes of the different $\Lb\to\jpsi \PLambda^*$ backgrounds, although we do not know their yields due to lack of knowledge of the relative $\PLambda^*\to\PSigma^0\pi^0$ branching fractions. We use separate shapes in the fit for the backgrounds corresponding to the $\PLambda(1405)$, $\PLambda(1520)$ and $\PLambda(1600)$ resonances. These backgrounds are simulated, processed through the event selections and fit using Gaussian kernel shapes.
We collectively model the sum of the remaining $\PLambda^*$ and $\psi(2S)$ backgrounds in the fit using a Gaussian shape. Note that our aim here is not to accurately disentangle each source of background, but only to model their collective sum.

A third background source arises from $\PXi_b\to\jpsi\PXi$ decays, where $\PXi\to\PLambda\pi$, when the pion from the $\PXi$ decay is not reconstructed. This background is modeled by a Gaussian kernel shape fit to simulated $\PXi_b^-\to\jpsi\PXi^-$ decays, which are partially reconstructed as $\jpsi\PLambda$. The normalization of this background is determined by fully reconstructing $\PXi_b^-\to\jpsi\PXi^-$ decays in data and simulation to obtain an efficiency-corrected yield. The reconstruction uses the criteria in Ref.~\cite{Aaij:2019ezy}. The reconstructed $\jpsi\PXi^-$ mass distribution in data is shown in the Supplemental material.
The efficiency-corrected yield is multiplied by the relative efficiency of reconstructing $\PXi_b^-\to\jpsi\PXi^-,$ as $\jpsi\PLambda$, and then more than doubled to account for $\PXi_b^0\to\jpsi\PXi^0$ decays. The production rates are unequal mostly because  the $\PXi'_b(5935)^0$ state is too light to decay into $\PXi_b^-\pi^+$ so it always decays into the $\PXi_b^0$ baryon \cite{Aaij:2014yka}. In addition, we incorporate the  production measurements of other excited $\Xib$ resonances \cite{Aaij:2016jnn} to determine the inclusive production ratio of  $\PXi_b^0 /\PXi_b^-=1.37\pm 0.09$, where the uncertainty arises mainly from the production fraction measurements of the excited states. We further corrected for the lifetime ratio $\tau_{\PXi_b^-}/\tau_{\PXi_b^0}=1.08\pm 0.04$ \cite{Aaij:2014lxa}. This normalization is introduced into the final fit as a Gaussian constraint, and done separately for Run 1 and Run 2 data, as the detection efficiencies differ.

The remaining background comes mostly from random combinations of real $\jpsi$ and $\PLambda$, which contribute both above and below the $\Lb\to\jpsi\PLambda$ mass peak.  This combinatoric background is modeled using an exponential function.

The Run 1 and Run 2 mass distribution data are fit simultaneously, using a binned extended maximum-likelihood fit, where the efficiency-corrected relative yields of the $\Lb\to\jpsi\PSigma^0$ signal, and those of the three $\Lb\to\jpsi\PLambda^*$ decays, with respect to the $\Lb\to\jpsi\PLambda$ signal, are constrained to be the same in the two data sets. We define
\begin{equation}\label{eq:main_R}
  \mathcal{R} \equiv \frac{|A_{1}|^2}{|A_{0}|^2}=\frac{{\cal{B}}(\Lb\to\jpsi\PSigma^0)}{{\cal{B}}(\Lb\to\jpsi\Lz)}\cdot {\Phi_\Lb} = \frac{N_{\Lb\to\jpsi\PSigma}}{N_{\Lb\to\jpsi\PLambda}} \cdot \frac{\epsilon_{\Lb\to\jpsi\PLambda}}{\epsilon_{\Lb\to\jpsi\PSigma}} \cdot{\Phi_\Lb},
\end{equation}
where $N_{\Lb\to\jpsi\PSigma}$ and $N_{\Lb\to\jpsi\PLambda}$ are the yields of the $\Lb\to\jpsi\PSigma$ and $\Lb\to\jpsi\PLambda$ decays; $\epsilon_{\Lb\to\jpsi\PSigma}$ and $\epsilon_{\Lb\to\jpsi\PLambda}$ are their respective efficiencies, as estimated from simulation; the phase space correction factor, $\Phi_\Lb$, is 1.058. The free parameters of interest in the fit are $\mathcal{R}$, $N_{\Lb\to\jpsi\PLambda}$, and $N_{\PXi_b^0\to\jpsi\PLambda}$; $N_{\Lb\to\jpsi\PSigma}$ can be calculated from these.
Systematic uncertainties are folded into the fit components as Gaussian constraints. These include uncertainties on the simulated ratios of efficiencies for the different \Lb final states  with respect to the $\jpsi\PLambda$ final state, which  range from 1.4 to 2.4\%. The uncertainty on the relative normalization of the $\PXi_b\to\jpsi\PXi$ background is estimated to be 12.1\% for Run~1 and 9.8\% for Run~2. This has contributions from the fit yield of the fully reconstructed $\PXi_b^-\to\jpsi\PXi^-$ decay, the reconstruction and efficiency of finding the $\PXi^-\to\PLambda\pi^-$ decay, and the  $\PXi_b^-/\PXi_b^0$ lifetime ratio.

The results of the fit are shown in Fig.~\ref{LOG_RUN2_COLOR}, and reported in Table~\ref{tab:fit_results}. The fitted value for $\mathcal{R}$, is consistent with zero. In Fig.~\ref{LOG_RUN2_COLOR}, we illustrate what this component would look like if observed at the upper limit on $\mathcal{R}$.
We do not quote the yields of the $\Lb\to\jpsi\PLambda^*$ decays as these are highly correlated.
\begin{table}[b]
\centering
\caption{Results from the fit to the $\jpsi\PLambda$ mass distribution. The fitted yields are indicated by $N$. Note $N_{\PXi_b \to \jpsi \PXi}$ indicates the sum of $\PXi_b^-$ and $\PXi_b^0$ decays.}
\vspace{0.2cm}
\begin{tabular}{lccc}
\hline\hline
Parameter & Shared value & Run 1 value & Run 2 value\\
\hline
$\mathcal{R}$ & $(0 \pm 5.3)\cdot 10^{-4}$ & -- & --\\
$N_{\Lb\to\jpsi\PLambda}$       & -- & $4417 \pm 66$ & $16~\!970 \pm 130$ \\
$N_{\PXi_b \to \jpsi \PXi}$     & -- & ~~\!$23.3 \pm 5.7$ & ~~$139.7 \pm  21.9$ \\
$N_{\PXi_b^0 \to \jpsi \PLambda}$ & -- & ~~~$6.2 \pm 3.0$ & ~~$17.8 \pm 5.1$ \\
\hline
\hline
\end{tabular}
\label{tab:fit_results}
\end{table}

\begin{figure}[t]
\begin{center}
 \includegraphics[width=0.7\textwidth]{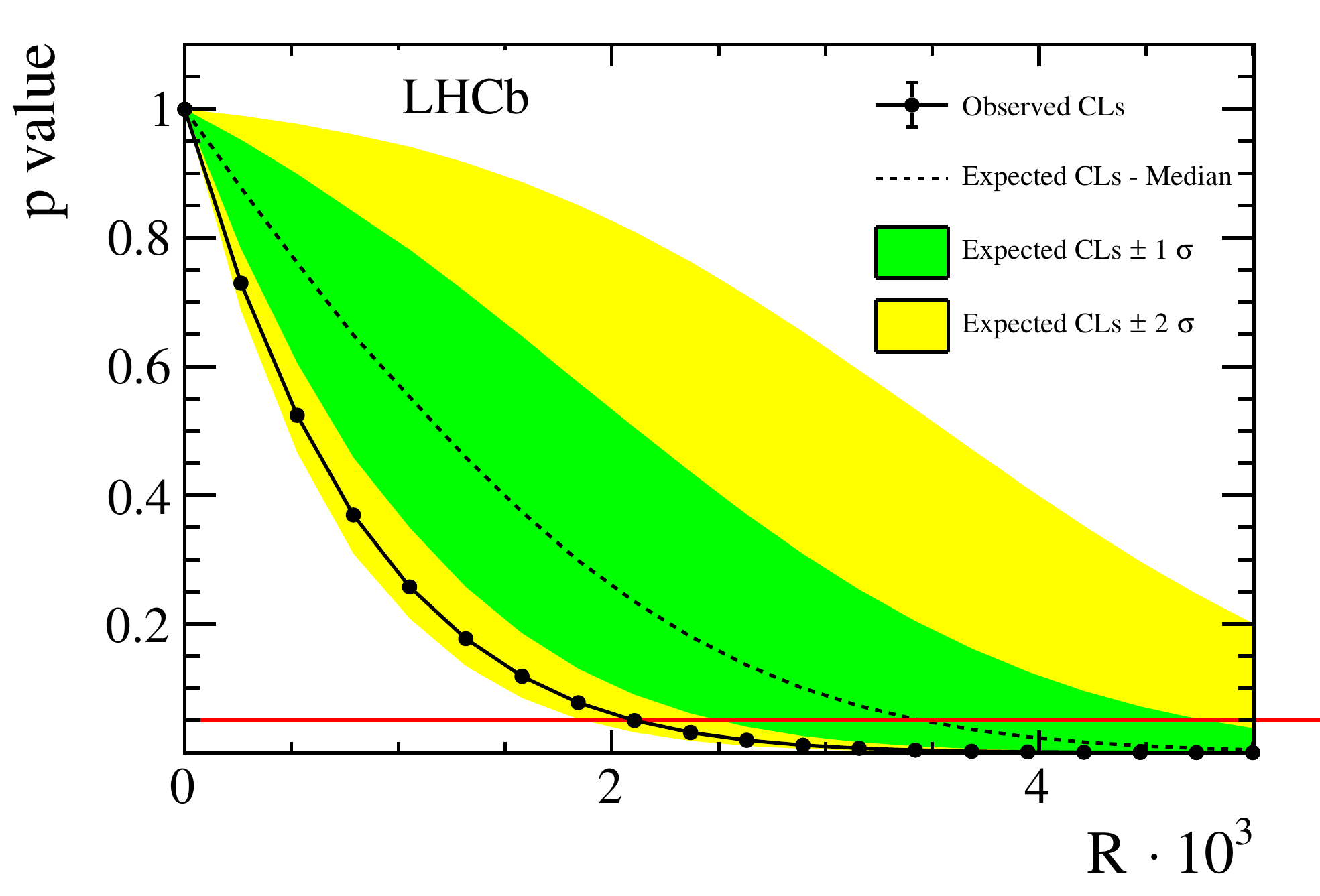}
\end{center}
\vskip -0.5cm
\caption{\small
Result of the hypothesis tests conducted using the CLs method by varying ${\cal{R}}$ is shown. The observed CLs distribution is shown by the round (black) points.  The expected CLs distribution (based on the background only hypothesis) is shown by the dashed line (black), with 1 and 2$\sigma$ uncertainty bands depicted in dark shaded (green) and light shaded (yellow) bands. The observed and expected upper limits are obtained by seeing where the bands cross the p-value of 0.05 shown as the horizontal (red) line.}
\label{fig:nominal_cls}
\end{figure}

To set an upper limit on ${\cal{R}}$ we use the CLs method \cite{Read:2002hq}.
The variation of the observed and expected CLs versus $\mathcal{R}$ is scanned from 0 to 0.005 and shown in Fig.~\ref{fig:nominal_cls}.  Our  observed upper limit on $\mathcal{R}$ is
\begin{equation}\label{eq:main_result}
  \mathcal{R} < 0.0021 \;\; \rm{at} \;\; 95\% \: CL. \nonumber
\end{equation}
Systematic uncertainties are incorporated in the fit and included in this limit. Further consistency checks include changing the fit range, eliminating the $\Lb\to\jpsi\PLambda^*$ background components one at a time, and fitting the $\Lb\to\jpsi\PLambda$ peak with different functions. These change the upper limit only by small amounts.

The Run 1 and Run 2 signal yields for $\PXi_b^{0} \to \jpsi \PLambda$ are listed in Table~\ref{tab:fit_results}. The statistical significance of the $\PXi_b^{0} \to \jpsi \PLambda$ signal is 5.6 standard deviations, obtained using Wilks' theorem \cite{wilks1938} and includes both the statistical and systematic uncertainties.
The branching fraction ratio ${\mathcal{B}(\PXi_b^{0} \to \jpsi \PLambda)}/{\mathcal{B}(\PXi_b^{0} \to \jpsi \PXi^{0})}$ is determined using the fully reconstructed $\PXi_b^- \to \jpsi\PXi^{-}$ sample described above.  To determine the branching fraction of $ \mathcal{B}(\PXi_b^{0} \to \jpsi \PXi^{0})$, we assume equal decay widths for the two different $\PXi_b\to \jpsi \PXi$ charge states and correct for the different neutral and charged $\PXi_b$ production rates as described above.
We use the measured lifetime ratio \cite{Aaij:2014lxa} to translate the decay width equality into the needed branching fraction.
The Run 1 and Run 2 results are consistent. Combining the two, we find
\begin{equation}
R_{\PXi_b}\equiv \frac{\mathcal{B}(\PXi_b^{0} \to \jpsi \PLambda)}{\mathcal{B}(\PXi_b^{0} \to \jpsi \PXi^{0})} = (8.2 \pm 2.1 \pm 0.9)\cdot 10^{-3},\nonumber
\end{equation}
where the first uncertainty is statistical the second is systematic, where the  leading source is the systematic uncertainty in the $\PXi_b^- \to \jpsi\PXi^{-}$ fit yield.

We convert $R_{\PXi_b}$ into a measurement of the amplitude ratio
\begin{equation}
\bigg|\frac{A_0}{A_{1/2}}\bigg|=\frac{1}{\lambda}\sqrt{\frac{R_{\PXi_b}}{\Phi_{\Xib}}}=0.37 \pm 0.06\pm 0.02\nonumber
\end{equation}
where $\Phi_{\Xib}=1.15$ is the relative phase space factor, and $\lambda=0.231$ is the relative Cabibbo suppression $|V_{cd}|/|V_{cs}|$, which is assumed equal to $|V_{us}|/|V_{ud}|$\cite{PDG}. Taking the $s$ and $u$ quarks in the $\PXi_b^0$ baryon to be a diquark state  with isospin 1/2 and combining with the null isospin of the $s$ quark from the $b$ quark decay, leads to isopsin 1/2 for the $\jpsi\PXi^{0}$ final state. On the other hand, for the Cabibbo suppressed transition with the isospin 1/2 $d$ quark, we have either isospin 0 or 1 final states. The former corresponds to $\jpsi\PLambda$, with the latter to $\jpsi\PSigma^{0}$, which we cannot currently measure. In order to predict the expected ratio of isospin amplitudes the SU(3) flavor \cite{Hiller:2013awa} $b$-baryon couplings must be taken into account \cite{Dery:2020lbc}. Then, if there are no other amplitudes, the theoretically predicted ratio corresponding to no preference between isospin 0 and 1/2 amplitudes
is $|{A_0}/{A_{1/2}}|$ equal to $1/\sqrt{6}$ $(\approx 0.41)$. Therefore, our result is consistent with no suppression of the isospin changing amplitude. These results are not precise enough to see the effects of SU(3) flavor symmetry breaking.

In conclusion, we set an upper limit in $\Lb\to\jpsi\PLambda(\PSigma^0)$ decays on the isospin amplitude ratio
\begin{equation}
|{A_{1}}/{A_{0}}|=\sqrt{\mathcal{R}}<{1}/{21.8}{\rm~ at~95\%~CL}.\nonumber
 \end{equation}
This limit is stringent and rules out isospin violation at a $\sim$1\% rate. Isospin violation has been seen at this level, for example, in $\rho-\omega$ mixing in $\Bzb\to\jpsi \pi^+\pi^-$ decays \cite{LHCb-PAPER-2014-058}. Our limit is consistent with the $\Lb$ being formed of a $b$ quark and a $ud$ diquark. This measurement also constrains non-Standard Model $A_1$ amplitudes contributing to $\Lb$ decays. Furthermore, our results support the quark model prediction of the $\Lb$ being an isosinglet.
Assumptions of isospin suppression in $\Lb \to \jpsi X$ decays made in past analyses are shown to be justified.
Finally, we report the discovery of the Cabibbo suppressed decay $\PXi_b^{0} \to \jpsi \PLambda$ and measure its branching fraction relative to $\PXi_b^{0} \to \jpsi \PXi^0$ to be $(8.2 \pm 2.1 \pm 0.9)\cdot 10^{-3}$. We see no evidence for the preference of either isospin amplitude in the ratio $|{A_0}/{A_{1/2}}|=0.37 \pm 0.06\pm 0.02$, as the prediction for the equality of isospin amplitudes is $1/\sqrt{6}$.

\section*{Acknowledgements}
%
%
\noindent We express our gratitude to our colleagues in the CERN
accelerator departments for the excellent performance of the LHC. We thank A. Ali, Y. Grossman, G. Isidori, Z. Ligeti, and J. Rosner for useful discussions. We thank the technical and administrative staff at the LHCb institutes. 
We acknowledge support from CERN and from the national agencies:
CAPES, CNPq, FAPERJ and FINEP (Brazil);
MOST and NSFC (China);
CNRS/IN2P3 (France);
BMBF, DFG and MPG (Germany);
INFN (Italy);
NWO (Netherlands);
MNiSW and NCN (Poland);
MEN/IFA (Romania);
MSHE (Russia);
MinECo (Spain);
SNSF and SER (Switzerland);
NASU (Ukraine);
STFC (United Kingdom);
DOE NP and NSF (USA).
We acknowledge the computing resources that are provided by CERN, IN2P3
(France), KIT and DESY (Germany), INFN (Italy), SURF (Netherlands),
PIC (Spain), GridPP (United Kingdom), RRCKI and Yandex
LLC (Russia), CSCS (Switzerland), IFIN-HH (Romania), CBPF (Brazil),
PL-GRID (Poland) and OSC (USA).
We are indebted to the communities behind the multiple open-source
software packages on which we depend.
Individual groups or members have received support from
AvH Foundation (Germany);
EPLANET, Marie Sk\l{}odowska-Curie Actions and ERC (European Union);
ANR, Labex P2IO and OCEVU, and R\'{e}gion Auvergne-Rh\^{o}ne-Alpes (France);
Key Research Program of Frontier Sciences of CAS, CAS PIFI, and the Thousand Talents Program (China);
RFBR, RSF and Yandex LLC (Russia);
GVA, XuntaGal and GENCAT (Spain);
the Royal Society
and the Leverhulme Trust (United Kingdom).

\newpage
\appendix
\section{Supplemental material}
The material here comprises the main BDT input variables in Section~\ref{sec:BDT}, the projection of the fit to the Run 1 $\jpsi\PLambda$ mass spectrum in Section~\ref{sec:Run1plot}, and the fits to the $\jpsi\PXi^-$ mass spectra in Section~\ref{sec:jpsiXi}.

\subsection{Final BDT variables}
\label{sec:BDT}
The variables used to train the BDT are:
\begin{enumerate}\setlength\itemsep{0.1em}%
  \item The $\chi^2$ of the global fit done by constraining the masses of the $\jpsi$ and $\PLambda$ particles to their known values and their momenta to the point of common origin.
  \item The minimum $\chi^2_{\rm IP}$ of the $\Lb$ candidate with respect to all the PVs.
  \item The DIRA of the $\Lb$ candidate with respect to its best PV, where DIRA of a particle is defined as its displacement vector with respect to the vertex.
  \item The significance of the flight distance of the $\Lb$ candidate with respect to the best PV.
  \item The minimum $\chi^2_{\rm IP}$ of the $\jpsi$ candidate with respect to all the PVs.
  \item The mass of the two muons forming the $\jpsi$ candidate.
  \item The significance of the $\PLambda$ candidates flight distance with respect to the \jpsi vertex.
  \item The DIRA of the $\PLambda$ candidate with respect to to the \jpsi vertex.
  \item The flight distance of the $\PLambda$ candidate with respect to to the \jpsi vertex.
  \item The DIRA of the $\PLambda$ candidate with respect to the best PV.
  \item The absolute difference between the mass of the $\PLambda$ candidate and its known value.
  \item The minimum $\chi^2_{\rm IP}$ of the $\PLambda$ candidate with respect to all the PVs.
  \item The minimum $\chi^2_{\rm IP}$ of the proton with respect to all the PVs.
  \item The probability that the proton from the $\PLambda$ candidate is not a real track, called a ``ghost." 
  \item The $p_\mathrm{T}$ of the proton from the $\PLambda$ decay.
  \item The particle identification of the proton from the $\PLambda$ decay.
  \item The ghost probability of the pion from the $\PLambda$ decay.
  \item The minimum $\chi^2_{\rm IP}$ of the pion from the $\PLambda$ decay with respect to all the PVs.
  \item The $p_\mathrm{T}$ of the pion from the $\PLambda$ decay.
  \item Output of the isolation BDT.
\end{enumerate}
\newpage
\subsection[Projection of the overall fit to the jpsiPLambda mass spectrum in the Run 1 data]{\boldmath Projection of the overall fit to the $\jpsi\PLambda$ mass spectrum in the Run 1 data}
\label{sec:Run1plot}

\begin{figure}[h]
\vskip -0.35cm
\centering
\includegraphics[width=0.9\textwidth]{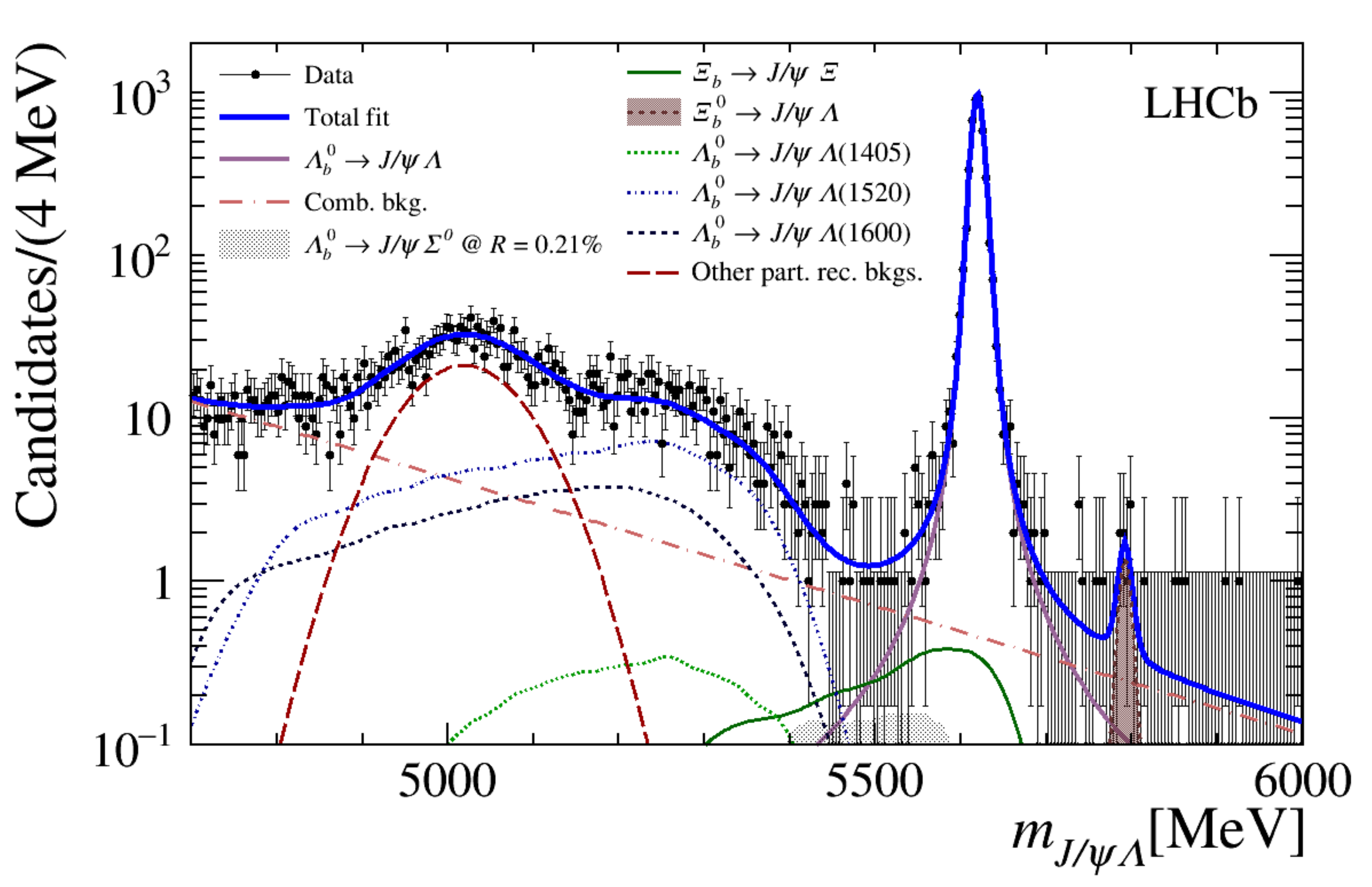}%
\caption{\small Distribution of $\jpsi\PLambda$ mass for Run 1 data. Error bars without data points  indicate empty bins. Also shown is the projection of the joint fit to the data. The thick (blue) solid curve shows the total fit. The $\Lb\to\jpsi\PSigma^0$ signal component is artificially scaled to its measured upper limit. The rest of the shapes are identified in the legend.}
\label{LOG_RUN1_COLOR}
\end{figure}

\subsection[Fits to the JpsiXi mass spectrum]{\boldmath Fits to the $\jpsi\Xires^-$ mass spectrum}
\label{sec:jpsiXi}
\begin{figure}[h]
\centering
\includegraphics[width=0.45\textwidth]{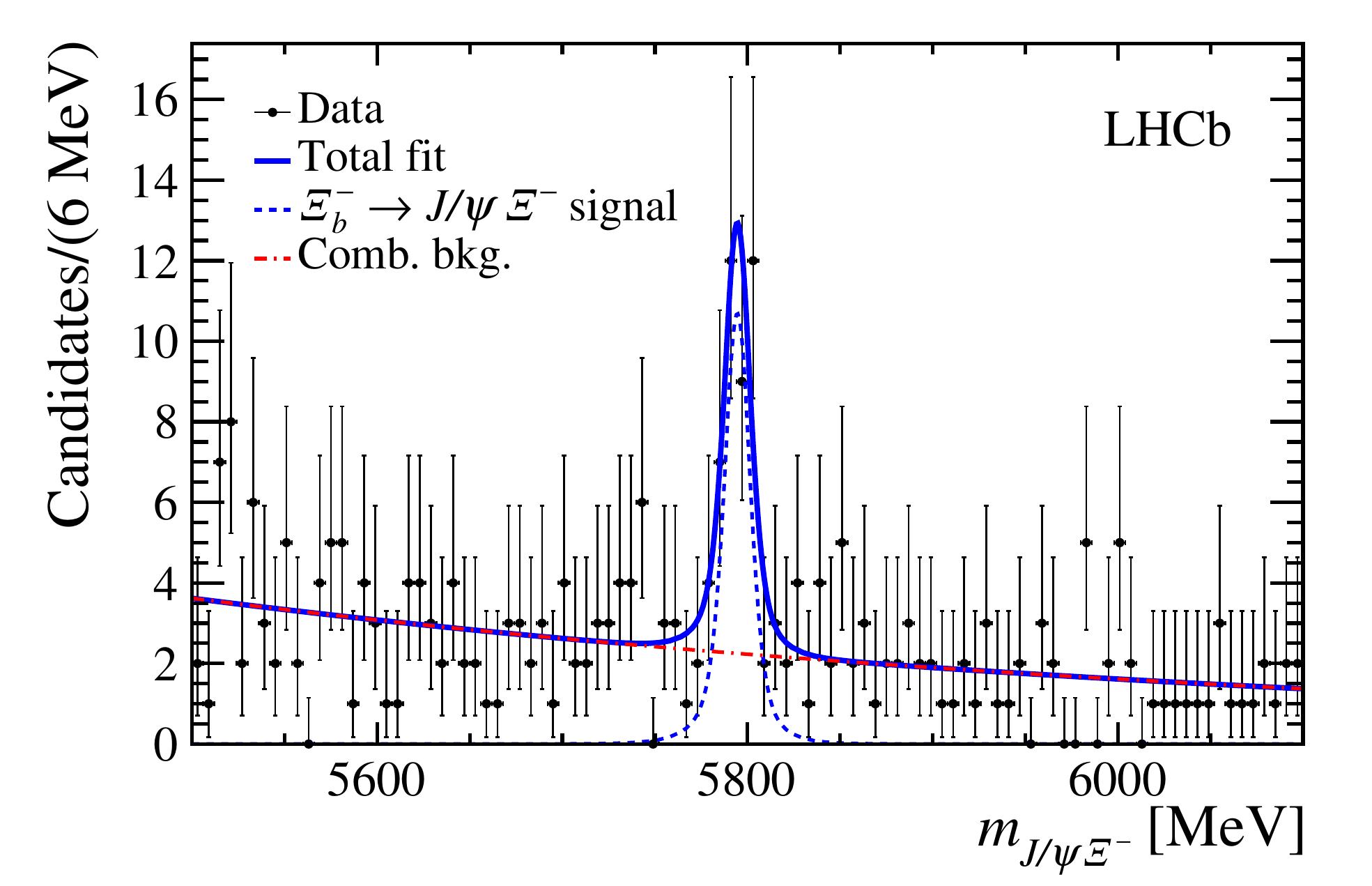}
\includegraphics[width=0.45\textwidth]{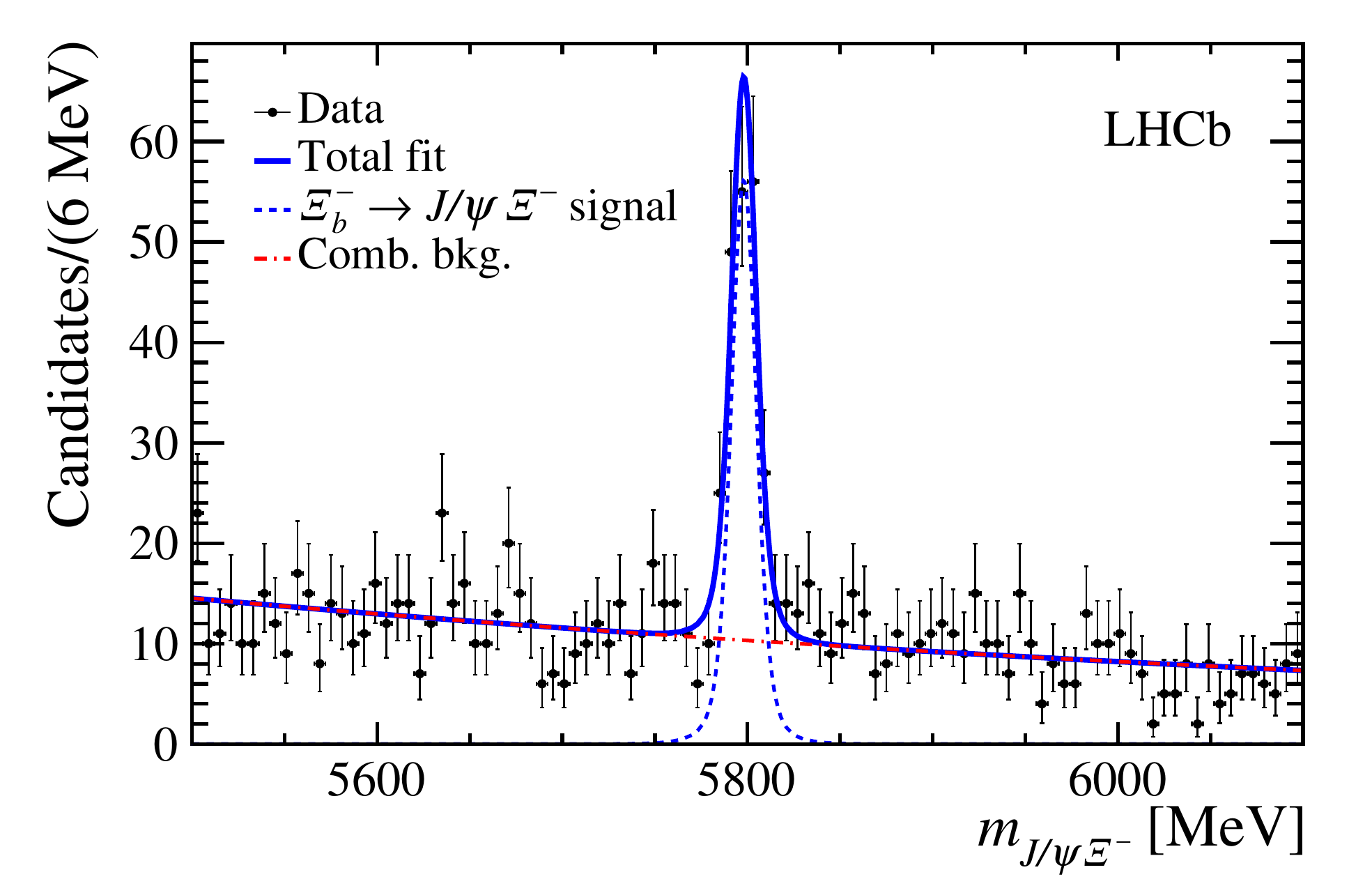}
\caption{\small Distributions of $\jpsi\PXi^-$ mass shown as points with error bars for (left) Run 1 and (right) Run 2 data.  The total fit to data is shown as a solid (blue) curve. The $\PXi_b^-\to\jpsi\PXi^-$ signal is fit with the sum of two Crystal Ball functions with the same mean and width, but different tail parameters, shown as the dashed (blue) curve. The combinatorial background shape is fit with an exponential function, shown as a dashed (red) curve.}
\label{Fit_JpsiXi_Data}
\end{figure}
\newpage
\newpage

\addcontentsline{toc}{section}{References}
\ifx\mcitethebibliography\mciteundefinedmacro
\PackageError{LHCb.bst}{mciteplus.sty has not been loaded}
{This bibstyle requires the use of the mciteplus package.}\fi
\providecommand{\href}[2]{#2}

\newpage
\centerline
{\large\bf LHCb collaboration}
\begin
{flushleft}
\small
R.~Aaij$^{31}$,
C.~Abell{\'a}n~Beteta$^{49}$,
T.~Ackernley$^{59}$,
B.~Adeva$^{45}$,
M.~Adinolfi$^{53}$,
H.~Afsharnia$^{9}$,
C.A.~Aidala$^{80}$,
S.~Aiola$^{25}$,
Z.~Ajaltouni$^{9}$,
S.~Akar$^{66}$
% ...

{\footnotesize \it

% ...
$ ^{9}$Universit{\'e} Clermont Auvergne, CNRS/IN2P3, LPC, Clermont-Ferrand, France\\
% ...
$ ^{25}$INFN Sezione di Milano, Milano, Italy\\
% ...
$ ^{31}$Nikhef National Institute for Subatomic Physics, Amsterdam, Netherlands\\
% ...
$ ^{45}$Instituto Galego de F{\'\i}sica de Altas Enerx{\'\i}as (IGFAE), Universidade de Santiago de Compostela, Santiago de Compostela, Spain\\
% ...
$ ^{49}$Physik-Institut, Universit{\"a}t Z{\"u}rich, Z{\"u}rich, Switzerland\\
% ...
$ ^{53}$H.H. Wills Physics Laboratory, University of Bristol, Bristol, United Kingdom\\
% ...
$ ^{59}$Oliver Lodge Laboratory, University of Liverpool, Liverpool, United Kingdom\\
% ...
$ ^{66}$Los Alamos National Laboratory (LANL), Los Alamos, United States\\
$ ^{67}$Syracuse University, Syracuse, NY, United States\\
% ...
$ ^{80}$University of Michigan, Ann Arbor, United States, associated to $^{67}$\\
\bigskip
% ...
}
\end{flushleft}

\end{document}